\documentclass[11pt,twoside]{article}
\usepackage{./asp2014}

\aspSuppressVolSlug
\resetcounters

\begin{document}

\title{Massive Stars: Some Open Questions and the role of Multi-Object Spectroscopy}
\author{A. Herrero$^{1,2}$
\affil{$^1$Instituto de Astrofisica de Canarias, La Laguna, Tenerife, Spain}
\affil{$^2$Universidad de La Laguna, La Laguna, Tenerife, Spain} \email{ahd@iac.es}}

\paperauthor{A. Herrero}{ahd@iac.es}{0000-0001-8768-2179}{Instituto de Astrofisica de Canarias}{}{La Laguna}{Tenerife}{E-38205}{Spain}

\begin{abstract}
Massive stars are key ingredients in the evolution of the Universe. Yet, important uncertainties and limits persist in our understanding of these objects, even in their early phases, limiting their application as tools to interpret the Universe. Here we review some of these open questions and argue that large samples are needed to answer them, both in the Milky Way and nearby galaxies. Multiobject spectroscopy plays a crucial role in this process. 
\end{abstract}

\section{Introduction}
By definition, massive stars end their lives with one of the most energetic phenomena in the Universe: a Supernova (SN) explosion (and presumibly, long-duration Gamma Ray Bursts sometimes). But massive stars are also related to other mighty phenomena along their whole life. They are extremely luminous objects that make their surroundings shine thanks to their intense UV fields, injecting huge amounts of mechanical energy through powerful stellar winds present from the early phases to the Red Supergiant (RSG) and Wolf-Rayet (WR) ones, including extreme stages with outbursts and eruptions, like Luminous Blue Variables (LBV). The matter yielded to the environment is often contaminated with nuclear processed material mixed from the inner layers, increasing the Helium and metals abundances in the Universe.

Massive stars have been proposed to be agents of the reionization of the Universe \citep[e.g.,][]{robertson10}. They have left their chemical footprint in early epoch stars \citep{aoki14}, opening a way to study the first moments of our Galaxy and are main sources behind the Star Formation Rate diagnostics \citep[SFR,][]{KE12}. These diagnostics indicate that the peak of SFR in the Universe happened at about a redshift z$\approx$ 2-3 \citep{bouwens12}. At that time, however, the properties of the Universe can be quite variable. Metallicity, for example, ranges from somewhat above solar down to about 10$^{-3}$ Z$_\odot$ in different objects \citep{cucchiara15}. Many objects have metallicities between 1.0 Z$_\odot$ and 0.1 Z$_\odot$, even at z$\approx$5.

It is well known that the main parameter determining the structure and evolution of a star is its mass. A second parameter is its chemical composition, often parametrized through its metallicity. Besides, mass-loss rate and angular momentum play an important role for massive stars (not so much for intermediate and low mass stars). Together, they determine the emergent flux that illuminates the material around, indirectly revealing in many cases the presence of the massive star. The combination of all these parameters determines the evolution of the star. Different scenarios have been proposed for the evolution of massive stars \citep{langer12, groh13}, without having still reached a global consensus. 

In the next sections, we review the status of our knowledge about the parameters determining the stellar evolution (mass, mass-loss rate and rotation at different metallicities) and the impact on their surroundings (emergent fluxes, or, as a proxy, the stellar effective temperature, T$_{\rm eff}$). We pay particular attention to the early stages, on or close to the Main Sequence, that we might naively think to be well known. We must also remark here the emergent phenomena that are getting increasing attention in the community: multiplicity, magnetic fields, atmospheric motions and stellar variability. They are not specifically treated here, but they are mentioned when appropiated.

\section{The stellar mass}
\cite{figer05} determined in his study of the Arches cluster that the upper mass limit, the highest mass a star reaches in Nature, is about 150 M$_\odot$. This number resulted from the study of the Present-Day Mass Function in the cluster, indicating that, if more massive stars were present, we should see the corresponding traces of previous SNe explosions. More recently, however, \cite{crowther10} found initial masses between 160 and 320 M$_\odot$ for stars a1, a2, a3 and c in the core of R136 in 30 Dor, with a1 being the record holder. They combined the UV-optical spectroscopic analysis with revised NIR photometry, and compared with evolutionary tracks they calculated. This means that the masses of Crowther et al. are model dependent (both from the point of view of atmosphere and evolutionary models) and subject to calibration issues. The high stellar density in the region means that binarity is an issue, although authors make a careful analysis trying to discard this possibility.

Of course, it is much better to use dynamical masses determined from binaries, particularly if they are eclipsing binaries. In Table 1 we see that the components in the most massive binaries known have masses slightly in excess of 100 M$_\odot$. However, these are present masses, and it is possible that the initial masses of these stars were higher. Very recently, \cite{sana13} have determined that the total present mass of the binary R144 in 30 Doradus is 200-300 M$_\odot$, estimating that the initial mass of the system could be as high as 400 M$_\odot$, which would mean that at least one of the components should have reached 200 M$_\odot$ at birth. We see that with these data the spectral type of the most massive stars is in the range WNh7-WNh5, while the mass range covers from 116 M$_\odot$ to some close to or even above 200 M$_\odot$.

However, most of the stars have to be analyzed as individual objects. Their masses have been obtained from two main techniques: (a) placing them in the Hertzsprung-Russell Diagram (HRD) and comparing with predictions of evolutionary tracks (the so-called {\it evolutionary masses}); and (b) by determining their stellar parameters from the spectra using model atmospheres and obtaining their radii and masses ({\it spectroscopic masses}). Although both techniques should give the comsistent masses, \citet{h92} showed that there is a systematic discrepancy between both (the {\it mass discrepancy}). Since then both, atmosphere and evolutionary models, have improved (particularly, with line-blanketing in model atmopsheres and rotational mixing in evolutionary ones). There have even been claims that the problem was solved thanks to these improvements \citep{WV10}, at least within errors. However, the literature is plaged with examples of the mass discrepancy. Discrepancies are found in all galaxies with accurate determinations of stellar parameters (Milky Way, LMC and SMC) and using different models. Recent examples are the works by \cite{martins12} and \cite{mahy15}. The first authors found spectroscopic masses that are lower than evolutionary ones for M$\le$ 25$\rm M_\odot$, and the latter find the same plus higher spectroscopic than evolutionary masses for M$\ge$ 40$\rm M_\odot$. This behaviour resembles that found by \cite{repolust04}, \cite{mokiem07a} and \cite{mokiem06} in the Milky Way, the LMC and the SMC. On the other hand, \cite{morrell14} found a systematic difference in three 30 Dor eclipsing binaries between the dynamical and evolutionary masses, although individual differences were within the errors.

{\it Thus we may conclude that the mass discrepancy is still controversial and remains as a problem, but at a milder scale (20-50$\%$) than when it was first detected.}
 
\begin{table}
\label{tabmass}
\caption{Most massive binaries known. R144 in 30 Dor \citep{sana13} could be even more massive (see text)}
\smallskip
\begin{center}
{\small
\begin{tabular}{lcrll}  
\tableline
\noalign{\smallskip}
System & Spectral Types & Mass (M$_\odot$) & Comments & References\\
\noalign{\smallskip}
\tableline
\noalign{\smallskip}
NGC 3603-A1 & WN6h+O & 116$\pm$31 &                      & \cite{schnurr08} \\
                       &                 &   89$\pm$16 &                      &                       \\
R 145             & WN6h+O  & 116$\pm$33 & Minimum masses & \cite{schnurr09} \\
                      &                 &  48$\pm$20 &                                &                     \\
WR21a          & WN+O      &  87              &                           & \cite{gamen07} \\
                     &                  &  87              &                           &                      \\
WR20a          & WNh6+WNh6 & 82.7      &                          & \cite{rauw05} \\
                     &                         & 81.9     &                          &                    \\
\noalign{\smallskip}
\tableline 
\noalign{\smallskip}
\tableline\
\end{tabular}
}
\end{center}
\end{table}

\section{The effective temperature scale}
The T$_{\rm eff}$ scale of O-type stars is used as a proxy for the emergent stellar fluxes, that determine the energy reemitted by the surroundings (either dust heated by these fluxes or emission from surrounding nebulosities). The scale has important implications for bolometric corrections, luminosities and ionizing fluxes, and is especially interesting for applications in star forming regions of galaxies. \cite{martins05} presented a temperature scale for O-stars in our Galaxy that has been extensively used as a reference. They collected data from the literature that were obtained in comparable ways and used state-of-the-art model atmospheres to present the temperature scale for O3-O9 stars in the Milky Way. Their temperature scale lowered previous scales by several thousand Kelvin, having a significant impact on all other physical magnitudes.

The temperature scale of O-stars has been extended to the Magellanic Clouds by \citet{mokiem06, mokiem07a} and \citet{massey09} (see also references therein). These authors find that, globally, the temperature scale in these galaxies is higher than in the Milky Way. A consistent result was obtained by \cite{GH13} in IC\,1613, where they found an even hotter scale as expected for lower metallicities (but see next section). All these temperature scales are based on the He ionization balance, which does not suffices for the hottest objects. More recently \cite{rivero12b} determined also the temperature scale for O-stars in the LMC, using the ionization balance of N, which allowed them to analyze earlier stars, in which the \ion{He}{i} lines were absent or very weak. They found agreement with previous temperature scales, but noted that for the earliest types a higher slope was needed (i.e., a more rapid increase of the temperature versus spectral type). \cite{sabin14} in their analysis of 30Dor stars in the VLT-FLAMES Tarantula Survey (VFTS) could not confirm (nor deny) this behaviour for O4-O2 types, as most of their earliest types were shown to be composite spectra. The comparison with the work of \cite{besten14} in 30 Dor is less straightforward because of differences in the analysis methodology, but the global behaviour is similar. Finally, we may cite the very recent work by \cite{camacho15} in the even lower-Z galaxy Sextans A. The large uncertainties in stellar parameters imposed by the faint spectra however precludes any firm conclusion.

The temperature scale of B-supergiants in the Milky Way and the LMC has been determined by \cite{MP08} and \cite{trundle07}. Their scales fit well with those from the previous authors.

We shall point here to the effects of small number statistics presented by \cite{ssimon14}. These authors have shown that the scatter in effective temperature at a given spectral type and luminosity class (thus, for exactly the same spectral classification) can be larger than the involved uncertainties. The reason is the discrete character of the spectral classification scheme combined with the change of stellar parameters during evolution, that compensate each other to produce the same ionization degree. For example, an O7V star of Galactic metallicity may have a T$_{\rm eff}$ between 36.0 and 38.5 kK and a gravity between 3.60 and 4.10 dex (note that they are correlated). The same ionization ratio for He can be produced with different combinations of T$_{\rm eff}$ and logg. This way, the slope for the Galactic temperature scale may vary depending on the specific stars chosen. Moreover, the the conclusions when comparing the slope of the temperature scale in two galaxies may be different for different samples. \cite{ssimon14} show two examples for the comparison of Galactic and SMC scales for the O-dwarfs in the Large Magellanic Cloud extracted from Monte-Carlo simulations. In the first one, the  temperature scale in the LMC is clearly hotter than in the Milky Way; in the second one, both are similar (see their Fig. 3). Therefore we must realize that the comparison of results between different galaxies is subject to these effects of low number statistics. This is more certain for galaxies beyond the Magellanic Clouds, where the data are really scarce.

{\it We conclude that the T$_{\rm eff}$ scale of O-stars is lower than 10 years ago, that shows a slight increase (heating) with decreasing metallicity and that the different sources agree with each other within uncertainties. We may also conclude that the data for B supergiants agree reasonably well with those of O-stars}. In the warning corner we shall include the fact that only two atmosphere codes have been used (FASTWIND and CMFGEN, see \cite{santolaya97, puls05, HM98}), with similar physical processes, and that the small number statistics may play a role for the limited samples available.

At the other side of the HRD, the effective temperature scale of Red Supergiants has also been subject to attention in the last years. \cite{levesque06} analyzed RSGs fitting TiO bands to updated MARCS models \citep{gustafsson03} and found hotter temperatures than previous analyses. This was good news, as it solved a consistency problem between the stellar temperatures previously determined from spectra and the evolutionary models. Moreover, recent analyses by \cite{meynet15} indicate that the radii derived from Levesque et al. agree well with evolutionary calculations at relatively low luminosities (log L/L$_\odot$ < 5.0) and with interferometric measurements at higher luminosity (but now departing from evolutionary predictions). However, \cite{davies13} fitting the Spectral Energy Distributions and using the Far Infrared Method obtain even higher temperatures than Massey et al. The discrepancy between both results can be attributed to the formation of TiO bands in upper layers, where opacity, structure and 3D effects are expected to be large. Of course, this raises again a problem with the evolutionary models, although Davies et al. indicate that the position of the T$_{\rm eff}$ limit in RSGs (i.e., the coolest T$_{\rm eff}$ RSGs can reach) depends strongly on the adopted overshooting parameter and is thus very unceratin. On the other hand, the results by Davies et al. (see their Section 5.5.3.) may help solving the question of the lack of high-mass SN progenitors from RSGs \citep{smartt09} (see next Section).

\section{The mass-loss}
Mass-loss in massive stars was recognized as an ubiquotous phenomenon after the launch of an UV spectrograph on board of an Aerobee rocket. \cite{morton67} observed strong UV P-Cygni profiles, indicative of strong mass-loss, in two spectroscopically normal OB supergiants, $\zeta$ Ori (O9.5 Ib) and $\epsilon$ Ori (B0 Ia). The importance of massive star winds was soon recognized and the fundamental theoretical frame was developed by \cite{cak75} (with improvements by \cite{ppk86} and \cite{FA86}). They showed that stellar winds in hot stars are driven by the absoprtion of radiation field momentum by metal lines. Soon, mass-loss was implemented in evolutionary calculations \citep[see][]{CM86}.

The fundamental prediction from the Radiatively Driven Wind (RDW) theory is a relation between the wind momentum gained by the escaping matter and the stellar luminosity. This is the so-called Wind Momentum - Luminosity Relationship (WLR, \cite{kudritzki95}):
$$
log D_{mom} = log(\dot{M}v_\infty R^{0.5})= ({{1}\over{\alpha^\prime}}) log L + const(Z, \rm{SpT})
$$
where $\dot{M}$ is the the stellar mass-loss, $v_\infty$ the wind terminal velocity, R and L the stellar radius and luminosity in solar untis, $\alpha^\prime$ a parameter that depends on the ratio of strong to weak lines in the wind acceleration zone (and thus on metallicity) and the constant at the end of the expression depends on the stellar metallicity and the spectral type.

This relationship has been confirmed for the range of metallicities between the Milky Way and the Magellanic Clouds by \cite{mokiem07b}, who obtained $\dot{M} \propto 0.72\pm0.15$ from the analysis of optical spectra. In Fig. 4 of Mokiem et al. we see that the observed slope of the relationship (${1}\over{\alpha^\prime}$) follows the theoretical expectations \citep{vink01} for the Milky Way, the LMC and the SMC, indicating that the basic physical mechanism is well understood. However, we identify two important problems in that figure: a vertical shift between the observed and theoretical relationships for all three galaxies and a limiting magnitude, below which the observed winds are weaker than theoretically expected. This limiting magnitude depends on the galaxy. Sometimes, when using UV resonance lines as diagnostics, mass-loss rates are orders of magnitude lower than predicted. We speak here usually of {\it weak winds}\footnote{A weak wind star should have a wind strength lower than stars of similar spectral class, not just lower than theoretically predicted; see \cite{walborn09}}.

Effects of wind clumping may be related to both points. These are density inhomogeneities in the wind that produce clumps with a higher density than the rest of the wind. The main observational diagnostics for mass-loss (and in particular H$_\alpha$, the one used by Mokiem et al.) rely on the emission through recombination processes, that depend on $\rho^2$. Therefore a clumped wind treated as a homogeneous one will result in an overestimation of its density (and thus its mass-loss rate) because $<\rho^2>$  is larger than $ <\rho>^2$. There are presently many evidences for the presence of wind clumping. The classical description of wind clumping considers small scale optically thin clumps in an almost void interclump medium. This is the so-called microclumping and it is subject to the effects on $\rho^2$ we just described. However, clumps may also be optically thick. In such case we speak of macroclumping, where effects of porosity in spatial or velocity spaces\footnote{Radiation scaping because of low opacity, either due to low density or to frequency shifts} may also play a role. Here, processes depending linearly on $\rho$ (like absorption in UV resonance lines) are affected and give mass-loss rates lower than actual ones. 

The classical diagnostic for mass-loss (the H$_\alpha$ line) is not only affected by clumping, but is also insufficient when trying to determine small mass-loss. Other diagnostics, like the radio and IR continua, or the Br$_\alpha$ line in the L-band are much more sensitive and help to dissentangle the spatial behaviour of clumping \citep[see][]{najarro11}. Thus, {\it multiwavelength observations are required to analyze mass-loss and clumping}. Mass-loss rates derived from bound-free absorption of X-ray line emission are particularly robust because they are less affected by clumping effects, whereas theoretical predictions \citep{vink01} overestimate actual mass-loss rates by a factor 2-3 (a nice discussion can be found in \cite{puls15}).

At metallicities beyond those of the Magellanic Clouds we face increasing problems. First, metals become less abundant, and their spectral signatures weaker; second, the distance to the interesting stars becomes larger, with two undesired effects: stars are fainter and crowding increases. Nevertheless, the reward is large, as we can test our theories at unprecedent low metallicities, approaching conditions in the early Universe.

Following theory, we would expect the winds in Z < Z$_{SMC}$ to be lower than in the SMC.  However, very metal poor Luminous Blue Variable stars
with strong optical P~Cygni profiles have been found in the Local Group 
\citep{h10} and in farther galaxies \citep[e.g.][]{drissen01,izotov11}. This not necessarily a problem for the RDW theory, as the mechanism behind the mass-loss in LBVs is not fully understood. More challenging, \cite{tramper14} (following \cite{tramper11}) report 10 stars with wind momenta stronger than predicted in IC\,1613, WLM and NGC\,3109,  Local Group galaxies with oxygen abundances corresponding to a metallicity $\rm \sim 1/7-1/10 \, Z_{\odot}$. This would indicate a breakdown of the RDW theory when the metallicity is very low (or alternatively, the emergence of another mechanism driving more efficiently the wind at very low metallicities, see \cite{lucy12}). However, Tramper et al. results are not confirmed by other researchers. \cite{h10} studied an Of star in IC\,1613 that could exhibit a strong wind, but \cite{h12} showed that the wind acceleration may be anomalous and its strength is reduced to SMC levels when the actual terminal velocity is used, and not a value scaled from the escape velocity, as it was done in \cite{h10}, \cite{tramper11} and \cite{tramper14}. Conclusive work has been presented by \cite{garcia14}, who obtained UV spectra of IC\,1613 stars and measured their terminal velocities. They presented strong evidence that the scaling of the terminal and escape velocities shows a large scatter. Moreover, they show that {\it he UV spectra of IC\,1613 stars display spectral signatures in the UV that indicate at least an SMC-like iron abundance}. This is in agreement with spectral analyses of M supergiant stars by \cite{taut07}. Also, \cite{hosek14} have shown a similar effect in NGC\,3109. Finally, recent work by \cite{bouret15} also shows smaller wind momenta for IC\,1613 stars than those obtained by Tramper et al., together with evidences in favour of an SMC-like iron abundance in this galaxy. Thus, Tramper et al. results may be explained by a combination of overestimated terminal velocities and an assumed metallicty derived from oxygen together with the assumption of a solar $\alpha$/Fe ratio (whereas the analyzed galaxies seem to have a subsolar $\alpha$/Fe ratio). Nevertheless, we must remember that we are in the realm of low number statistics: the number of analyzed objects is still worringly small. And uncertainties are large because of the difficulties mentioned in the above paragraph. {\it More analyses of objects in low metallicity galaxies are needed, particularly in galaxies with independent evidence of a solar $\alpha$/Fe ratio}.

Mass-loss rates are poorly known for RSGs and usually empirical, highly uncertain formulae are used \citep[see][]{smith14}. However, mass-loss rate during this phase has a strong impact on the subsequent evolution. \cite{meynet15} have shown that mass-loss during this phase determines the presence or absence of blue loops. Although this mass-loss should be much larger than presently assumed, its effects could explain the lack of massive progenitors of core collapse SNe. \cite{smartt09}, comparing observations to standard evolutionary calculations, conclude that the initial masses of post-RSG core collapse SNe progenitors range between 8$^{1.0}_{-1.5}$ and 16.5$\pm{1.5}$ M$_\odot$, clearly below the expected value of 25 M$_\odot$. With enhanced mass-loss rates, the initial masses obtained for RSGs exploding as SNe would be higher, thus reconciling observations and models (see also the end of next section). Alternatively, may be that stars more massive than about 20 M$_\odot$ suffer a dark core collapse, i.e., form directly a black hole without SN explosion (or with a very faint event, see \cite{smartt09}).

\section{Rotation}
Angular momentum is the physical magnitude that determines the structure and evolution of stars, besides mass and mass-loss. The special interest on rotation in massive stars was innitiated because of the evidences for mixing processes acting in the stellar interior. These evidences, known for more than twenty years, rely on abundance anomalies in some stars, mainly (but not only) enhanced He abundances and CNO abundances that reflect the effects of the nuclear reactions in the CNO cycle \citep[see][and references therein]{MM00}. Recent works confirm these evidences in stars in the Milky Way \cite[f.e.,][]{martins12, martins15} and the Magellanic Clouds (\cite{mokiem06}, \cite{rivero12a}). Theoretical models with induced rotational mixing have been proposed to explain the observed abundance patterns (see \cite{langer12, MM00}). In these models, rotation mixes material from the interior towards the surface, thus reproducing the enhancements of He and the CNO patterns observed at the surface. The efficiency of the process increases with stellar mass and with rotational rate: the higher the mass and the faster the rotation, the earlier and more intense will be the contamination by CNO products. A significant initial rotational velocity is needed to start the mixing, and if it is high enough, evolution proceeds homogeneously: the star has a constant abundance profile and it evolves towards the left from the Zero Age Main Sequence. Of course, the effect may be time dependent. As the star evolves, it losses mass and angular momentum while expanding and the rotational speed decreases (at least at the surface, but note that the important factor is the angular momentum in the interior; rigid body rotation or a given angular velocity profile with depth link the angular momenta at the surface and in the interior).

The difficulties of the rotational mixing began with the work presented by \cite{hunter08}. These authors represented the N abundance versus the projected rotational velocity for 135 early B-stars in the LMC. They found that although many stars followed the expected correlation between vsini and N enhancement, there were at least two groups in the diagram that could not be explained by rotational mixing: group 1 contained stars with large rotation and low or no N enhancement; group 2 contained young (high gravity) stars with low rotational velocities and large N enhancements. The number of stars in both groups makes highly improbable that they are due to projection or low-number statistics
effects. 

A promising alternative scenario for the mixing processes is the evolution in binary (or multiple) systems (for a description of the effects of binarity, see the review by A. de Koter in these proceedings). We will only indicate here that group 1 could for example be explained by systems in which a more massive companion tranfers mass and angular momentum to the gainer, that is thus accelerated. If the acceleration has happened recently, the rotational mixing has not had time to act. Group 2 could be explained by systems in which the mass transferred was already N rich (because of a strong stellar wind, or because of rotational mixing in the primary). Alternatively (or complementarily) magnetic fields, either fossile or generated during the binary interaction, may brake down the star, although the presence of magnetic fields in massive O-stars seems to be modest (around 8$\%$, see \cite{wade14}, \cite{fossati15}). The large parameter space of massive binary systems allows also other possibilities, including SN explosions (that could result in a fast rotating runaway, after the secondary suffers a kick) and stellar merges (resulting in fast rotating, rejuvenated single stars). 
 
 In a very recent work, \cite{martins15} compare the CNO abundances from their analyses of 74 stars with the predictions of rotating Geneva evolutionary tracks, concluding that 80$\%$ of the objects can be explained by rotational mixing. However, we shall note that the main observational fact is the presence at the surface of material that has been processed by the CNO cycle, without an easy way to separate the different scenarios. {\it Thus we conclude that nuclear processed material is brought to the surface of early massive stars, but the exact process responsible for it is not clearly identified (as all of them will produce a similar pattern)} and we have to look for second order differences (like correlations between parameters or  time scales). While rotational mixing and binary interaction can probably account qualititively for all cases, other processes like magnetism and inner instabilities have still to be addressed in a consistent way.
 
A key parameter in these studies is the rotational velocity. We must emphasize here that we measure the present, projected rotational velocity, whereas what we would like to know is the initial, unprojected rotational velocity. Nevertheless, the first contains important information and is what is directly available to the observation. The most recent extensive determination of rotational velocities of O stars in the Milky Way has been carried out by \cite{SH14} (see there for a list of other efforts and a description of techniques). These authors have determined the rotational velocity of nearly 200 Galactic stars including {\it macroturbulence} (see below). They find a bimodal distribution for vsini (already seen by \cite{CE77}). The main peak lies at 40-60 km$^{-1}$, although we have to take into account that there is a limit in the lowest rotational velocity that can be determined, depending on spectral resolution, S/N ratio and assumed broadening mechanisms (if we neglect some mechanism, its effect will automatically be assigned to rotation -- or the last mechanism we deal with). This limit will be different for different stars, and runs in the study of \cite{SH14} from 20 to 40 km s$^{-1}$. The secondary peak lies at about 300 km s$^{-1}$ and could correspond to a population of stars that have been accelerated in the course of their evolution in a binary system, as suggested by \cite{demink13}. In fact, their relative number in the sample of  Sim\'on-D\'iaz \& Herrero (23\% of the stars rotate with projected rotational velocities in excess of 200 km s$^{-1}$) agrees remarkably well with the predictions of de Mink et al. (20$^{+5}_{-10}$ \%). The maximum rotational velocity measured by Sim\'on-D\'iaz \& Herrero is 430 km s$^{-1}$. 

The global distribution for Galactic O stars presented by Sim\'on-D\'iaz \& Herrero has the same characteristics as the one presented by \cite{ramirezag13} in 30 Doradus, only with slightly different values for the main characteristics. The main peak appears in 30 Dor at 40-80 km s$^{-1}$ (with a lower spectral resolution, however), the secondary one appears slightly above 400 km s$^{-1}$ and the highest velocity is 620 km s$^{-1}$. {\it We note that although the highest velocities found in the MW and specially in the the LMC are sufficient to expect homogeneous evolution, there are still no firm evidences supporting it. Moreover, in spite of the differences found in the numbers by \cite{SH14} and \cite{ramirezag13}, both distributions are very similar, although one is a sample of bright, nearby stars belonging to many regions in the Milky Way and the other is a sample of stars belonging to the same star forming region in the LMC}. These results are consistent with those by other authors, even in the SMC \citep[see f.e.][]{mokiem06, PG09}.

However, we must be careful when determining rotational velocities, as they are obtained from the broadening of the spectral lines and these are affected by other broadening mechanisms. Besides the well known mechanisms that we find in textbooks (like the natural, Doppler or the different collisional broadening mechanisms) spectral lines in massive O and B stars are affected by a brodening mechanism of still unknown origin. It is usually called {\it macroturbulence}, and it is parametrized in the same way as in cool stars, with a radial-tangential mechanism, but it has nothing to do with hydrodynamical turbulence \cite[see][and references therein]{gray02, SH14}. Still, it has to be taken into account when deriving the rotational velocity, specially when this is below 120 km s$^{-1}$ \cite{SH14}. Additional difficulties may be present when vsini is low \citep{SH14}, particularly when magnetic fields are present \citep[see][]{sundqvist13}.

This extra broadening can be quite informative. It has been proposed that it may be linked to pulsations (kappa-driven gravity mode pulsations, see \cite{aerts09}, \cite{ssimon10} and references therein). However, the most recent results indicate that this may not be the solution for many massive stars. \cite{ssimon15a} presented the position in the HRD of stars dominated by macroturbulent broadening. The upper part of the diagram is fully populated with such stars. Moreover, he showed that the positions of the stars do not completely overlap with the predicted instability domains corresponding to high order g-mode pulsations originated by the kappa-mechanism and that macroturbulence may be large close to the Zero Age Main Sequence, where no instability is expected. {\it Thus this kind of pulsations do not seem to explain the presence of macroturbulent broadening in high-mass stars, although the evidence is much more supportive for stars with masses below 20-25 M$_\odot$}. Recently, \cite{grassitelli15} propose that turbulent pressure, originated by sub-surface convective layers, may be at the root of this extra broadening and \cite{AR15} conclude that it may be explained by convection-driven internal gravity waves (IGW), based on 2D simulations (although in a lower, 3 M$_\odot$ model, modified such that it might represent the IGW spectrum of a 30 M$_\odot$ star).  {\it All these works indicate that macroturbulent broadening may be the key for studying the inner zones below the stellar surface}.

Rotation may also play an important role in massive star evolution at the other edge of the HRD. Surface abundances of RSGs are higly sensitive to rotation in that phase \citep{meynet15}. This may help (together with the application of the Ledoux criterion, see \cite{georgy14}) to reconcile the contradictory findings by \cite{saio12} that some blue supergiants have pulsational properties in agreement with post-RSGs objects, but abundance enhancements that are much smaller than predicted by evolutionary models. The enhanced mass-loss rates during the RSG phase commented in the previous section would be responsible for the existence of the blue loops, and the enhanced rotation would be responsible for the lower abundances. However, we have to remark that the agreement is by far not complete. The observed abundances of post-RSG candidates (i.e., B-supergiants having pulsational properties expected for post-RSG stellar structures) are still clearly smaller than those predicted by the non-standard evolutionary calculations. And there are still no observational evidences supporting the presence of two different populations among B-supergiants (as it would be the case if post-RSG blue loops do exist). Rotation may help to clarify the situation. Post-RSG B-supergiants should have much lower rotational velocities than their pre-RSG counterparts. However, this would imply to be able to determine accurate rotational velocities at very low levels, at or below 5 km s$^{-1}$.

\section{Massive stars and Multi-Object Spectroscopy}

The open issues presented above can only be solved using a wide base of stellar data. Spectra contain the largest amount of information, and thus it is important to construct spectroscopic databases, with wide wavelength coverage. Multiepoch information is particularly important, not only to get information about multiplicity and interaction in binaries, but also to dissentangle the roles of the different physical processes, from magnetic fields to atmospheric motions and wind variability. The success of programs like The FLAMES Survey of Massive Stars \citep{evans05}, the VLT-FLAMES Tarantula Survey \citep{evans11} or other spectroscopic surveys of OB stars in the Milky Way like GOSSS \citep{sota11}, IACOB \citep{ssimon15b} and OWN \citep{barba10} or those including the search for magnetic fields like MiMeS \citep{wade14}, BinaMIcS \citep{alecian15} and BOB \citep{morel15}, shows the great potential of this approach. The large amount of objects required for these databases makes the use of spectrographs with large multiplexing capabilities unavoidable. WEAVE, the future multiobject spectrograph for the WHT, has characteristics that will allow us to explore the population of massive stars in the Milky Way with unprecedent detail, being able to reach the most luminous stars in nearby galaxies. These data will nicely complement Gaia, helping to the effort of a detailed map of the Milky Way. Together with the suite of multiobject spectrograph already in use in large telescopes (like f.e., FLAMES, MUSE and KMOS at VLT, OSIRIS at GTC, DEIMOS and MOSFIRE at Keck) and foressen for the nearby future (like MEGARA, EMIR and MIRADAS at GTC) it will provide a large collection of spectra to study the physics of massive stars and apply it to the study of the entire Universe.

\acknowledgements I want to thank the organizers of the Conference for the invitation to give this keynote talk and their patience to receive the manuscript. Thanks also to F. Najarro, J. Puls and S. Sim\'on-D\'iaz for the reading of the manuscript, and inspiring comments and discussions. This work has been supported by Spanish MINECO under grants AYA2010-21697-C05-04, AYA2012-39364-C02-01 and SEV 2011-0187-01.

\begin{thebibliography}{}
\bibitem[Aerts et al.(2009)]{aerts09} Aerts, C., Puls, J., Godart, M. \& Dupret, M.-A. \ 2009, \aap, 508, 409
\bibitem[Aerts \& Rogers(2015)]{AR15} Aerts, C. \& Rogers, T.M. \ 2015, \apjl, 806, L33
\bibitem[Alecian et al.(2015)]{alecian15} Alecian, E., Neiner, C., Wade, G.A. \ 2015, in {\it New windows on massive stars}, IAUS 307, 330
\bibitem[Aoki et al.(2014)]{aoki14} Aoki, W., Tominaga, N., Beers, T.C., Honda, S. \& Lee, Y.S. \ 2014, Science, 345, 912  

\bibitem[Barb\'a et al.(2010)]{barba10} Barb\'a, R.H., Gamen, R., Arias, J.I. et al. \ 2010, RevMexAA, 38, 30
\bibitem[Bestenlehner et al.(2014)]{besten14} Bestenlehner, J.M., Gr\"afener, G., Vink, J.S. et al. \ 2014, \aap, 570, A38
\bibitem[Bouwens et al.(2012)]{bouwens12} Bouwens, R.J., Illingworth, G.D., Oesch, P.A. et al. \ 2012, \apj, 754, 83 
\bibitem[Bouret et al.(2015)]{bouret15} Bouret, J.-C., Lanz, T., Hillier, D.~J., et al.\ 2015, \mnras, 449, 1545 

\bibitem[Camacho et al.(2015)]{camacho15} Camacho, I., Garcia, M., Herrero, A. \& Sim\'on-D\'iaz, S. \ 2015, \aap, in press
\bibitem[Castor, Abbott \& Klein(1975)]{cak75} Castor, J.I., Abbott, D.C. \& Klein, R.I. \ 1975, \apj, 170, 325
\bibitem[Chiosi \& Maeder(1986)]{CM86} Chiosi, C. \& Maeder, A. \ 1986, ARA\&A, 24, 329
\bibitem[Conti \& Ebbets(1977)]{CE77} Conti, P.S. \& Ebbets, D. \ 1977, \apj, 213, 438
\bibitem[Crowther et al.(2010)]{crowther10} Crowther, P.~A., Schnurr, O., Hirschi, R. et al. \ 2010, \mnras, 408, 731 
\bibitem[Cucchiara et al.(2015)]{cucchiara15} Cucchiara, A., Fumagalli, M., Rafelski, M. et al. \ 2015, \apj, 804, 51

\bibitem[Davies et al.(2013)]{davies13} Davies, B., Kudritzki, R.P., Plez, B. et al. \ 2013, \apj, 767, 3
\bibitem[de Mink et al.(2013)]{demink13} de Mink, S.E., Langer, N., Izzard, R.G., Sana, H. \& de Koter, A. \ 2013, apj, 764, 166
\bibitem[Drissen et al.(2001)]{drissen01} Drissen, L., Crowther, P.A., Smith, L.J. et al., 2001, \apj, 546, 484

\bibitem[Evans et al.(2005)]{evans05} Evans, C.J., Smartt, S.J., Lee, J.-K. et al., 2005, A\&A, 437, 467  
\bibitem[Evans et al.(2011)]{evans11} Evans, C.J., Taylor, W.D., H\'enault-Brunet, V. et al., 2011, A\&A, 530, A108 

\bibitem[Figer(2005)]{figer05} Figer, D.F. \ 2005, Nature, 434, 192
\bibitem[Fossati et al.(2015)]{fossati15} Fossati, L., Castro, N., Schoeller, M. et al. \ 2015, \aap, in press
\bibitem[Friend \& Abbott(1986)]{FA86} Friend, D.B. \& Abbott, D.C. \ 1986, \apj, 311, 701

\bibitem[Gamen et al.(2007)]{gamen07} Gamen, R., Barb\'a, R. Morrell, N. et al. \ 2007, BAAA, 50, 105
\bibitem[Garcia \& Herrero(2013)]{GH13} Garcia, M., \& Herrero, A.\ 2013, \aap, 551, A74 
\bibitem[Garcia et al.(2014)]{garcia14} Garcia, M., Herrero, A., Najarro, F., Lennon, D.J., \& Urbaneja, M.A.\ 2014, \apj, 788, 64
\bibitem[Georgy et al.(2014)]{georgy14} Georgy, C., Saio, H. \& Meynet, G. \ 2014, MNRAS 439, L6
\bibitem[Grassitelli et al.(2015)]{grassitelli15} Grassitelli, L., Fossati, L., Sim\'on-D\'iaz, S. et al. \ 2015, \apjl, 808, L31
\bibitem[Gray (2002)]{gray02} Gray, D.F. \ 2002, Observation and analysis of stellar atmospheres, John Wiley \& Sons
\bibitem[Groh, Meynet \& Ekstr\"om (2013)]{groh13} Groh, J.H., Meynet, G. \& Ekstr\"om, S. \ 2013, \aap, 550, L7
\bibitem[Gustafsson et al.(2003)]{gustafsson03} Gustafsson, B. Edvarsson, B., Eriksson, K. et al. \ 2003, in {\it Stellar Atmosphere Modelling}, ASP Conf. Series 288, I. Hubeny, D. Mihalas, K. Werner eds., p. 331

\bibitem[Herrero et al.(1992)]{h92} Herrero, A., Kudritzki, R.P., Vilchez, J.M. et al., 1992, A\&A 261, 209
\bibitem[Herrero et al.(2010)]{h10} Herrero, A., Garcia, M., Uytterhoeven, K. et al., 2010, A\&A, 513, A70
\bibitem[Herrero et al.(2012)]{h12} Herrero, A., Garcia, M., Puls, J., et al.\ 2012, \aap, 543, A85 
\bibitem[Hillier \& Miller(1998)]{HM98} Hillier, D.~J., \& Miller, D.~L.\ 1998, \apj, 496, 407 
\bibitem[Hosek al.(2014)]{hosek14} Hosek, M. W. Jr, et al. 2014, \apj, in preparation
\bibitem[Hunter et al.(2008)]{hunter08} Hunter, I., Brott, I.,  Lennon, D.J. et al., 2008, \apj, 676, L29

\bibitem[Izotov et al.(2011)]{izotov11} Izotov, Y.I., Guseva, N.G., T.X., Fricke, K.J., \& Henkel, C.\ 2011, \aap, 533, A25

\bibitem[Kennicutt \& Evans (2012)]{KE12} Kennicutt, R.C. Jr. \& Evans, J.J. II \ 2012, ARA\&A, 50, 531 
\bibitem[Kudritzki, Lennon, \& Puls(1995)]{kudritzki95} Kudritzki, R.-P., Lennon, D.~J., \& Puls, J.\ 1995,{\it Science with the VLT}, p246 

\bibitem[Langer(2012)]{langer12} Langer, N. \ 2012, ARA\&A, 50, 107
\bibitem[Levesque et al.(2006)]{levesque06} Levesque, E.M., Massey, P., Olsen, K.A.G. et al. \ 2006, \apj, 645, 1102
\bibitem[Lucy(2012)]{lucy12} Lucy, L.~B.\ 2012, \aap, 543, A18  

\bibitem[Maeder \& Meynet(2000)]{MM00} Maeder, A., \& Meynet, G.\ 2000, ARA\&A, 38, 143 
\bibitem[Mahy et al.(2015)]{mahy15} Mahy, L., Rauw, G., De Becker, M., Eenens, P. \& Flores, C.A. \ 2015, \aap, 577, A23
\bibitem[Markova \& Puls(2008)]{MP08} Markova, N., \& Puls, J.\ 2008, \aap, 478, 823 
\bibitem[Martins et al.(2005)]{martins05} Martins, F., Schaerer, D., Hillier, D.~J. \ 2005, \aap, 436, 1049 
\bibitem[Martins et al.(2012)]{martins12} Martins, F., Mahy, L., Hillier, D.J. \& Rauw, G. \ 2012, \aap, 538, A39
\bibitem[Martins et al.(2015)]{martins15} Martins, F., Herv\'e, A., Bouret, J.-C. et al. \ 2015, \aap, 575, A34
\bibitem[Massey et al.(2009)]{massey09} Massey, P., Zangari, A.M., Morrell, N.I. et al. \ 2009, \apj, 692, 618
\bibitem[Meynet et al.(2015)]{meynet15} Meynet, G., Chomienne, V., Ekstr\"om, S. et al. \ 2015, \aap, 575, A60
\bibitem[Mokiem et al.(2006)]{mokiem06} Mokiem, M.~R., de Koter, A., Evans, C.~J., et al.\ 2006, \aap, 456, 1131   
\bibitem[Mokiem et al.(2007a)]{mokiem07a} Mokiem, M.R., de Koter, A., Evans, C.J. et al., 2007, A\&A, 465, 1003   
\bibitem[Mokiem et al.(2007b)]{mokiem07b} Mokiem, M.~R., de Koter, A., Vink, J.~S., et al.\ 2007b, \aap, 473, 603  
\bibitem[Morel et al.(2015)]{morel15} Morel, T., Castro, N., Fossati, L. \ 2015, in {\it New windows on massive stars}, IAUS 307, 342 
\bibitem[Morrell et al.(2014)]{morrell14} Morrell, N.I., Massey, P., Neugent, K.F., Penny, L.R. \& Gies, D.R. \ 2014, \apj, 789, 139
\bibitem[Morton (1967)]{morton67} Morton, D.C. \ 1967, \apj, 147, 1017

\bibitem[Najarro, Hanson \& Puls(2011)]{najarro11} Najarro, F., Hanson, M.M. \& Puls, J. \ 2011, \aap, 535, 32


\bibitem[Pauldrach, Puls \& Kudritzki(1986)]{ppk86} Pauldrach, A., Puls, J. \& Kudritzki, R.P., 1986, 164, 86
\bibitem[Penny \& Gies(2009)]{PG09} Penny, L.R. \& Gies, D.R. \ 2009, \apj, 700, 844
\bibitem[Puls et al.(2005)]{puls05} Puls, J., Urbaneja, M.A., Venero, R. et al. \ 2005, \aap, 435, 669
\bibitem[Puls et al.(2015)]{puls15} Puls, J., Sundqvist, J.O. \& Markova, N. \ 2015, in New windows on massive stars, IAUS 307, 25

\bibitem[Ram\'irez-Agudelo et al.(2013)]{ramirezag13} Ram\'irez-Agudelo, O.H., Sim\'on-D\'iaz, S., Sana, H. et al. \ 2013, \aap, 560, A29
\bibitem[Rauw et al.(2005)]{rauw05} Rauw, G., Crowther, P.A., De Becker, M. \ 2005, \aap, 432, 985
\bibitem[Repolust et al.(2004)]{repolust04} Repolust, T., Puls, J. \& Herrero, A., 2004, A\&A, 415, 349
\bibitem[Rivero Gonz{\'a}lez et al.(2012a)]{rivero12a} Rivero Gonz{\'a}lez, J.~G., Puls, J., Najarro, F., \& Brott, I.\ 2012a, \aap, 537, A79  
\bibitem[Rivero Gonz{\'a}lez et al.(2012b)]{rivero12b} Rivero Gonz{\'a}lez, J.~G., Puls, J., Massey, P., \& Najarro, F.\ 2012b, \aap, 543, A95 
\bibitem[Robertson et al.(2010)]{robertson10} Robertson, B.~E., Ellis, R.~S., Dunlop, J.~S., et al.\ 2010, Nature, 468, 49

\bibitem[Sab\'in-Sanjuli\'an et al.(2014)]{sabin14} Sab\'in-Sanjuli\'an, C., Sim\'on-D\'iaz, S., Herrero, A. et al. \ 2014, \aap, 564, A39
\bibitem[Saio et al.(2012)]{saio12} Saio, H., Georgy, C. \& Meynet, G. \ 2012, MNRAS 433, 1246
\bibitem[Sana et al.(2013)]{sana13} Sana, H., van Boeckel, T., Tramper, F. et al. \ 2013, \mnras, 432, L26
\bibitem[Santolaya-Rey et al.(1997)]{santolaya97} Santolaya-Rey, A.E., Puls, J., Herrero, A. \ 1997, \aap, 323, 488
\bibitem[Schnurr et al.(2008)]{schnurr08} Schnurr, O., Casoli, J., Chene, A.N. et al \ 2008, \mnras, 389, L38
\bibitem[Schnurr et al.(2009)]{schnurr09} Schnurr, O., Moffat, A.F.J., Villar-Sbaffi, A. et al \ 2008, \mnras, 395, 823
\bibitem[Sim\'on-D\'iaz(2015)]{ssimon15a} Sim\'on-D\'iaz, S. \ 2015, in {\it New windows on massive stars}, IAUS 307, 194
\bibitem[Sim\'on-D\'iaz et al.(2015)]{ssimon15b} Sim\'on-D\'iaz, S., Negueruela, I., Ma\'iz-Apell\'aniz, J. et al. \ 2015, in {\it Highlights of Spanish Astronomy VIII}, 576
\bibitem[Sim\'on-D\'iaz et al.(2010)]{ssimon10} Sim\'on-D\'iaz, S., Herrero, A., Uytterhoeven, K. et al. \ 2010, \apjl, 720, L174
\bibitem[Sim\'on-D\'iaz \& Herrero(2014)]{SH14} Sim\'on-D\'iaz, S. \& Herrero, A. \ 2014, \aap, 582, 135 
\bibitem[Sim\'on-D\'iaz et al.(2014)]{ssimon14} Sim\'on-D\'iaz, S., Herrero, A., Sab\'in-Sanjuli\'an, C. et al. \ 2014, \aap, 570, L6  
\bibitem[Smartt et al.(2009)]{smartt09} Smartt, S.J., Eldridge, J.J., Crockett, R.M. \& Maund, J.R. \ 2009, MNRAS 398, 1409
\bibitem[Smith(2014)]{smith14} Smith, N. \ 2014, ARA\&A, 52, 487
\bibitem[Sota et al.(2011)]{sota11} Sota, A., Ma\'iz-Apell\'aniz, J., Walborn, N.R. et al. \ 2011, \apjs, 193, 24
\bibitem[Sundqvist et al. (2013)]{sundqvist13} Sundqvist, J., Sim\'on-D\'iaz, S., Puls, J. \& Markova, N. \ 2013, \aap, 559, L10 

\bibitem[Tautvai{\v s}ien{\.e} et al.(2007)]{taut07} Tautvai{\v s}ien{\.e}, G., Geisler, D., Wallerstein, G. et al. \ 2007, \aj, 134, 2318 
\bibitem[Tramper et al.(2011)]{tramper11} Tramper, F., Sana, H., de Koter, A. \& Kaper, L. \ 2011, \apjl, 741, L8
\bibitem[Tramper et al.(2014)]{tramper14} Tramper, F., Sana, H., de Koter, A., Kaper, L. \& Ram\'irez-Agudelo, O.H. \ 2014, \aap, 572, A36
\bibitem[Trundle et al.(2007)]{trundle07} Trundle, C., Dufton, P.L., Hunter, I. et al. \ 2007, \aap, 471, 625  


\bibitem[Vink, de Koter \& Lamers(2001)]{vink01} Vink, J.~S., de Koter, A. \& Lamers, H.J.G.L.M.\ 2001, \aap, 369, 574 

\bibitem[Wade et al.(2014)]{wade14} Wade, G.A., Grunhut, J., Alecian, E. et al. \ 2014, Proc. of the IAUS 302, {\it Magnetic fields through stellar evolution}, 265 
\bibitem[Walborn(2009)]{walborn09} Walborn, N.R. \ 2009, in {\it Massive Stars: from Pop. III and GRBs to the Milky Way}, eds. M. Livio \& E. Villaver, STScI Symp. Ser. 20, 167
\bibitem[Weidner \& Vink(2010)]{WV10} Weidner, C. \& Vink, J. \ 2010, \aap, 524, A98
\end{thebibliography}


\end{document}